\documentclass[aps,prl,reprint,longbibliography,superscriptaddress,floatfix,nobalancelastpage]{revtex4-2}
\usepackage{symbols}
\usepackage{xfrac}
\usepackage{amsfonts}
\usepackage{dsfont}
\usepackage{amsmath}
\usepackage{amssymb}
\usepackage{amsthm}
\usepackage{mathtools}
\usepackage{graphicx}
\usepackage[caption=false]{subfig}
\usepackage{floatrow}
\usepackage{color}
\usepackage{bm}
\usepackage[normalem]{ulem}
\usepackage[hidelinks]{hyperref}
\hypersetup{
     colorlinks   = true,
     citecolor    = blue,
     linkcolor    = blue
}
\usepackage{mathptmx}
\usepackage{enumitem}

\usepackage{relsize}
\usepackage{xcolor}
\usepackage{scalerel}
\usepackage{tikz}
\usetikzlibrary{svg.path}
\usepackage{soul}
\definecolor{orcidlogocol}{HTML}{A6CE39}
\tikzset{
  orcidlogo/.pic={
    \fill[orcidlogocol] svg{M256,128c0,70.7-57.3,128-128,128C57.3,256,0,198.7,0,128C0,57.3,57.3,0,128,0C198.7,0,256,57.3,256,128z};
    \fill[white] svg{M86.3,186.2H70.9V79.1h15.4v48.4V186.2z}
                 svg{M108.9,79.1h41.6c39.6,0,57,28.3,57,53.6c0,27.5-21.5,53.6-56.8,53.6h-41.8V79.1z M124.3,172.4h24.5c34.9,0,42.9-26.5,42.9-39.7c0-21.5-13.7-39.7-43.7-39.7h-23.7V172.4z}
                 svg{M88.7,56.8c0,5.5-4.5,10.1-10.1,10.1c-5.6,0-10.1-4.6-10.1-10.1c0-5.6,4.5-10.1,10.1-10.1C84.2,46.7,88.7,51.3,88.7,56.8z};
  }
}
\newcommand\orcid[1]{\href{https://orcid.org/#1}{\mbox{\scalerel*{
\begin{tikzpicture}[yscale=-1,transform shape]
\pic{orcidlogo};
\end{tikzpicture}
\ 
}{|}}}}

\newcommand{\red}[1]{{\color{red} {#1}}}
\newcommand{\stkout}[1]{\ifmmode\red{\text{\sout{\ensuremath{#1}}}}\else \red{\sout{#1}}\fi}
\newcommand{\figpanel}[2]{\hyperref[#1]{\ref*{#1}(#2)}}
\begin{document}

\title{Certification Time for Weak-Force-Mediated Entanglement between Levitated Particles}

\author{Yao-Tong Chen \orcid{0000-0003-3687-6541}}
\affiliation{ICFO – Institut de Ciencies Fotoniques, The Barcelona Institute of Science and Technology, Castelldefels, Barcelona 08860, Spain}

\author{Salambô Dago \orcid{0000-0002-1964-2375}}
\affiliation{University of Vienna, Faculty of Physics, Vienna Center for Quantum Science and Technology (VCQ), Boltzmanngasse 5, Vienna, 1090, Austria}
\affiliation{Institut de Sciences et d’Ingénierie Supramoléculaires, University of Strasbourg, France}

\author{Oriol Romero-Isart \orcid{0000-0003-4006-3391}}
\affiliation{ICFO – Institut de Ciencies Fotoniques, The Barcelona Institute of Science and Technology, Castelldefels, Barcelona 08860, Spain}
\affiliation{ICREA – Institucio Catalana de Recerca i Estudis Avançats, Barcelona 08010, Spain}

\author{Andreu Riera-Campeny \orcid{0000-0003-3260-993X}}
\affiliation{ICFO – Institut de Ciencies Fotoniques, The Barcelona Institute of Science and Technology, Castelldefels, Barcelona 08860, Spain}

\date{\today}

\begin{abstract}
Certifying motional entanglement generated by weak forces between levitated nanoparticles requires distinguishing entangled from separable states using finite, noisy measurement records. We show that the number of measurements required to certify Gaussian entanglement can vary among equally entangled states, and quantify this variation. We derive an attainable lower bound showing that, near the separability threshold, the required number of measurements diverges inversely with the square of the distance from this threshold. We translate the required number of measurements into an experimental certification time, comparing two protocols under comparable experimental conditions: release–recapture, in which particles evolve freely before measurement, and continuous monitoring, in which the steady state is measured continuously. We find that release–recapture enables entanglement certification at interaction strengths several orders of magnitude weaker. Moreover, minimizing certification time can favor states with less entanglement, revealing a trade-off between entanglement generation and its certification. These results highlight the role of certification in the design and optimization of ongoing efforts to probe weak-force-mediated entanglement between levitated nanoparticles.
\end{abstract}

\maketitle

Levitated mechanical systems provide a platform for exploring quantum phenomena at mesoscopic mass scales, with applications ranging from quantum sensing to tests of fundamental physics~\cite{Aspelmeyer2014, Moore2014, Moore2021, Afek2022, Carney2021, Aggarwal2022}. In levitated optomechanics~\cite{Millen2020, Gonzalez-Ballestero2021}, isolation in high vacuum and tunable optical~\cite{Rossi2025, Dago2024} and non-optical trapping potentials~\cite{Bonvin2024, Tomassi2026} have enabled advances in the control and cooling of both center-of-mass motion~\cite{Tebbenjohanns2021, Magrini2021, Ranfagni2022, Kamba2022, Kamba2025, Rossi2025} and rotational degrees of freedom~\cite{Kamba2023, Dania2024, Troyer2026}. A key next step is to extend this level of quantum control to interacting multipartite systems and, in particular, to experimentally demonstrate entanglement between spatially separated mechanical objects~\cite{Weiss2021, Carlon2025, Winkler2025}. The main challenge is to engineer interactions strong enough to generate entanglement while preserving the resulting quantum correlations against noise and decoherence~\cite{Aspelmeyer2022}. 

Generating entanglement, however, is only part of the challenge: it must also be certified from finite, noisy measurement data~\cite{BlumeKohout2010, Arrazola2013, Chevalier2020, Tilly2021, Cieslinski2024}. Entanglement certification is a statistical inference problem, in which measurements are used to estimate an entanglement witness, either directly or through state reconstruction~\cite{Genoni2008, Guhne2009}. Statistical uncertainty limits the ability to distinguish entangled from separable states, particularly near the separability boundary, where increasingly many measurements are required. In mechanical experiments, where independent realizations are acquired sequentially, this requirement translates into experimental time and demands environmental and system stability throughout the acquisition. The relevant question is therefore not only whether entanglement can be generated, but how long it takes to certify it with a prescribed statistical confidence.

In this work, we determine the number of measurements required to certify Gaussian entanglement from noisy measurements of the covariance matrix. We derive an attainable lower bound that diverges as the inverse square of the distance to separability, with a prefactor determined by the statistical fluctuations of the measured observables rather than by the entanglement strength alone. We then investigate the implications for weak-force-mediated entanglement between levitated nanoparticles, considering two experimental strategies. In release–recapture, the trapping potentials are switched off to allow free expansion and subsequently restored for measurement. Single-particle implementations of this protocol have recently been demonstrated~\cite{Steiner2025, Mattana2026}. In continuous monitoring, the particles remain trapped while their steady-state motion is measured, as in standard free-space optical-tweezer experiments~\cite{Magrini2021, Tebbenjohanns2021}. We find that the release duration minimizing certification time does not coincide in general with that maximizing entanglement, and that optimizing for certification rather than entanglement generation can reduce the required measurement time. Under comparable experimental conditions, release–recapture also enables certification at much weaker interaction strengths than continuous monitoring.

We consider the one-dimensional transverse motion of two identical massive particles, labeled $s\in\{\sysa,\sysb\}$, with dimensionless canonical quadratures satisfying $[\posop_s,\momop_{s'}]=2i\delta_{ss'}$. The particles interact through the potential
\begin{align}
    \hat V = \frac{\hbar g}{2}(\posop_\sysa - \posop_\sysb)^2,
    \label{eq:interaction}
\end{align}
obtained by expanding an attractive central potential to quadratic order in the transverse displacement~\cite{ThisSM}. This approximation requires the relative transverse displacement to remain small compared with the equilibrium separation $d$. To connect the coupling $g$ to physical interactions, we evaluate representative strengths for Coulomb~\cite{Frimmer2017}, Casimir--Polder~\cite{Emig2007, Weiss2021}, and gravitational forces~\cite{Lindner2005, Bose2017, Marletto2017, Aspelmeyer2014} between identical levitated spheres, using the parameters listed in Table~\ref{tab:physical_params} of the End Matter. In the release--recapture protocol, particles initially cooled with trapping frequency $\trapfreq$ are released freely to evolve freely under their mutual interaction before being recaptured for readout~\cite{Steiner2025, Mattana2026}. In continuous monitoring, the particles remain trapped while their positions are continuously measured and linear feedback is applied~\cite{Magrini2021, Tebbenjohanns2021}. In both protocols, we seek to certify entanglement of the unconditional two-particle state. We model environmental noise through momentum diffusion at rate $\phr_\dark$, with an additional measurement backaction at rate $\phr_\backaction$ when optical readout is active. Neither process introduces position diffusion. 

For Gaussian states, the covariance matrix $\cov$ contains all information relevant to entanglement~\cite{Weedbrook2012,Serafini2017}. Its elements are $\cov_{jk}=\av{\{\hat\psvec_j-\av{\hat\psvec_j},\hat\psvec_k-\av{\hat \psvec_k}\}}/2$, where we have defined the operator vector $\hat{\psvec} =(\posop_\sysa, \momop_\sysa, \posop_\sysb, \momop_\sysb)^\transpose$. To reconstruct $\cov$, we consider measurements that provide noisy estimates of the position and momentum quadratures simultaneously. Each measurement yields a Gaussian outcome $\psvec_n$ with covariance $\bbSigma = \cov + \cov_\meas$, where $\cov_\meas$ accounts for the added measurement noise. Such measurements can, for instance, be implemented by retrodiction after recapture~\cite{Lammers2024, Rossi2025}. For known zero means and calibrated measurement noise, an unbiased estimator from $\nm$ independent measurements is
\begin{align}
    \estimate{\cov} = \frac{1}{\nm} \sum_{n=1}^{\nm} \psvec_n\psvec_n^\transpose-\cov_\meas,
\label{eq:covariance_estimator}
\end{align}
Subtracting $\cov_\meas$ makes the covariance estimator unbiased, but does not remove the contribution of measurement noise to its statistical uncertainty, which depends on the total outcome covariance $\bbSigma=\cov+\cov_\meas$. Throughout, we assume perfectly calibrated measurement noise, neglecting calibration uncertainties that must be accounted for in a specific experimental implementation.

To determine the number of measurements required for entanglement certification, we consider a witness $\witness=f(\cov)$ with separability threshold $\witness_{\rm sep}$~\cite{Guhne2009}. We estimate the witness from the reconstructed covariance matrix as $\estimate{\witness} = f(\estimate{\cov})$, where the check denotes an estimate obtained from measurement data. Under suitable regularity conditions, this estimator is consistent and asymptotically normal~\cite{vanDerVaart1998, Casella2002}. We quantify the required resources using the $\nsigma$-sigma criterion
\begin{align}
    \witness + \nsigma\Std(\estimate{\witness}) < \witness_{\rm sep},
\label{eq:witness_certification}
\end{align}
where $\Std(x)=\sqrt{\expect[(x-\expect(x))^2]}$ and $\expect$ denotes averaging over measurement data. Here, we use the true witness value to determine the number of measurements required, whereas experimental certification would need the estimated witness for certification. For $N\gg1$, the witness estimator is asymptotically normal, with a bias negligible compared with its statistical uncertainty. We calculate these fluctuations analytically and determine the number of independent measurements required for certification.

To determine the statistical uncertainty of the entanglement witness, we propagate fluctuations in the estimated covariance matrix to the witness using the delta method~\cite{vanDerVaart1998, Casella2002}. For sufficiently large data sets and differentiable $f(\cov)$, this propagation is linear to leading order
\begin{align}
    \estimate{\witness} \simeq \witness + \trc[\matG(\cov) (\estimate{\cov}-\cov)], 
\label{eq:delta_method_expansion}
\end{align}
where the gradient matrix $\matG_{ij}(\cov) \equiv [\partial f / \partial \cov_{ij}](\cov)$, quantifies the sensitivity of the witness to covariance errors. The statistical errors in different elements of the estimated covariance matrix $\estimate{\cov}$ are generally correlated. Accounting for these correlations, the variance of the witness estimator is~\cite{vanDerVaart1998}
\begin{align}
    \Var(\estimate{\witness}) \simeq
    \sum_{ijkl} \matG_{ij}(\cov)
    \Cov(\estimate{\cov}_{ij}, \estimate{\cov}_{kl})
    \matG_{kl}(\cov),
\label{eq:general_variance_witness}
\end{align}
where $\Cov(x,y)=\expect[(x-\expect(x))(y-\expect(y))]$.

For any two-mode Gaussian state, entanglement is fully characterized by the Peres--Horodecki criterion~\cite{Peres1996, Horodecki1996, Simon2000}. The state is entangled if and only if the smallest symplectic eigenvalue $\sse$ of the partially transposed covariance matrix satisfies $\sse<1$. We therefore take $\witness\equiv\sse$, with
\begin{align}
\label{eq:tilde_nu_minus_general}
    \sse=f(\cov) = \sqrt{\frac{1}{2}\left[\tilde{\Delta}-\sqrt{\tilde{\Delta}^2-4\det(\cov)}\right]},
\end{align}
where $\tilde{\Delta} \equiv \det(\matA)+\det(\matB)-2\det(\matC)$, and with the block decomposition $\cov=(\matA, \matC; \matC^\transpose, \matB)$, where the semicolon separates block rows.

Applying Eq.~\eqref{eq:general_variance_witness} to the witness $\witness=\sse$, we obtain the asymptotic fluctuations of the witness estimator~\cite{ThisSM}
\begin{align}
    \Std[\estimate{\tilde{\nu}}_-] \simeq \sqrt{\frac{2\trc[\matG(\cov) \bbSigma\matG(\cov) \bbSigma]}{\nm}} \equiv \frac{\alpha}{\sqrt{\nm}}.
    \label{eq:std_deviation_sse}
\end{align}
For $\sse<1$, requiring the entanglement gap $1-\sse$ to exceed $\nsigma$ standard deviations yields the asymptotic number of measurements needed for certification
\begin{align}
\nm_\star\simeq\left(\frac{\nsigma\alpha}{1-\sse}\right)^2.
    \label{eq:Nstar}
\end{align}
Certification depends on both the entanglement gap $1-\sse$ and the fluctuation coefficient $\alpha$. The statistical fluctuations inherent to covariance estimation do not vanish at separability, making the inverse-square increase in the required number of measurements unavoidable, as shown below. For instance, changing $\sse$ only in the third decimal place, from $0.999$ to $0.990$, reduces the required number of measurements by two orders of magnitude at fixed $\alpha$, substantially easing experimental certification.

\begin{figure}
    \centering
    \includegraphics[width=\linewidth]{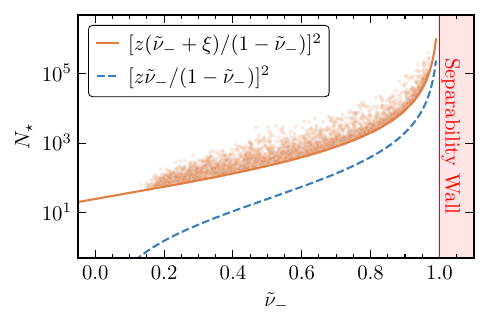}
    \caption{Sample complexity $\nm_\star$ for entanglement certification with $\nsigma=5$ and isotropic readout noise $\measinp =1$. The orange solid line shows the attainable bound $\alpha\geq\sse+\measinp$, while the blue dashed line represents the unphysical noise-free benchmark $\measinp =0$. The points correspond to $4000$ randomly sampled two-mode Gaussian states with symplectic eigenvalues $\lambda_k$ drawn uniformly from $[1,1.6]$. The covariance matrix is constructed as $\cov=\matS{\rm diag}(\lambda_1, \lambda_1, \lambda_2, \lambda_2)\matS^\transpose$, where $\matS$ consists of two-mode squeezing with parameter uniformly sampled from $[0,1]$, followed by local squeezing and rotations with parameters independently sampled for each mode from $[-1/2,1/2]$ and $[0,\pi]$, respectively, and a beam-splitter transformation with mixing angle uniformly sampled from $[0,\pi/2]$. Only states with $0.02<\tilde{\nu}_{-}<0.98$ are shown.}
    \label{fig:wall}
\end{figure}

In levitated optomechanics, continuous position measurements after recapture provide a record of the trapped motion. This record can be used to retrodict both position and momentum at the recapture time, effectively realizing a simultaneous noisy measurement of both quadratures~\cite{Lammers2024, Rossi2025}. We model the added measurement noise as isotropic, $\cov_\meas = \measinp\matI$ with $\measinp \geq1$, where $\measinp =1$ corresponds to the quantum-limited joint quadrature measurement. Under these conditions, we obtain~\cite{ThisSM}
\begin{align}
    \alpha^2 = \sse^2 + 2 \measinp\sse\trc[\matG(\cov)] + 2 \measinp^2 \trc[\matG(\cov)^2].
    \label{eq:measurement_penalty}
\end{align}
The first term depends only on $\sse$, whereas the terms arising from measurement noise also depend on the state through $\matG(\cov)$. As shown in the End Matter, Eq.~\eqref{eq:measurement_penalty} yields the attainable universal lower bound $\alpha\geq\sse+\measinp$, and hence
\begin{align}
    \nm_\star \gtrsim \left[\frac{\nsigma(\sse+\measinp)}{1-\sse}\right]^2.
    \label{eq:universal_std}
\end{align}
This bound is attained by symmetric thermal two-mode squeezed states. However, states with identical entanglement can require arbitrarily more measurements: local squeezing preserves $\sse$ while allowing $\alpha$ to grow without bound~\cite{ThisSM}. Thus, stronger entanglement lowers the minimum achievable cost but does not by itself guarantee more efficient certification. Figure~\ref{fig:wall} illustrates this effect, showing the number of measurements required for states with the same $\sse$, and hence the same entanglement. The attainable bound defines the lower envelope, while the unphysical noise-free limit ($\measinp=0$) provides a reference for the intrinsic sampling fluctuations of the quantum state, without additional readout noise. The spread above the bound reflects the dependence on the state's covariance matrix. As the separability threshold is approached, even the minimum number of measurements diverges as $(1-\sse)^{-2}$, imposing a fundamental statistical barrier to the experimental certification of weak entanglement.

We now apply the certification framework to the release--recapture protocol, starting from two independent thermal states with mean occupation $\nbar$. The collective quadratures $\posop_\pm=(\posop_\sysa\pm\posop_\sysb)/\sqrt{2}$ and $\momop_\pm=(\momop_\sysa\pm\momop_\sysb)/\sqrt{2}$ decouple the dynamics during release. The common mode evolves freely, while the differential mode oscillates at frequency $\freq_-=\sqrt{4g\trapfreq}$~\cite{ThisSM}. Their different evolution generates interparticle correlations. During release, the particles are subject only to dark momentum diffusion at rate $\phr_\dark$. After recapture, optical readout introduces additional backaction at rate $\phr_\backaction$. We assume negligible interaction between the particles after recapture and measurement and diffusion rates much smaller than $\trapfreq$. Under these conditions, retrodiction of a sufficiently long measurement record yields approximately isotropic added noise, $\cov_\meas\simeq\measinp\matI$~\cite{Lammers2024}, with
\begin{align}
\measinp \simeq \sqrt{\frac{\phr_\backaction+\phr_\dark}{\phr_\meas}},
\end{align}
where $\phr_\meas$ is the measurement rate. In the following, we use the detection efficiency $\efficiency_\detection\equiv\phr_\meas/\phr_\backaction$ to characterize the information acquired relative to measurement backaction.

To examine the weak-force limit, we consider $\epsilon=\sqrt{4g/\trapfreq}\ll1$ at a fixed ratio $r=\phr_\dark/g>0$, preserving the competition between dark diffusion and the entangling interaction. We also fix the interaction phase $\theta=\freq_-\tint$, which requires increasing the release duration as $\tint=\theta/(\epsilon\trapfreq)$ for weaker interactions. At fixed initial occupation $\nbar$, the entanglement witness approaches~\cite{ThisSM}
\begin{align}
    \sse\simeq rF(\theta),
    \label{eq:rr_sse_asymptotics}
\end{align}
where $F(\theta)$ is a dimensionless function of order unity. Thus, at fixed $r$ and $\theta$, the entanglement witness becomes independent of $\epsilon$ to leading order. In contrast, the fluctuation coefficient scales as
\begin{align}
    \alpha\simeq\frac{\measinp H(\theta)}{\epsilon},
    \label{eq:rr_am_asymptotics}
\end{align}
where $H(\theta)$ is also of order unity. Both functions $F(\theta)$ and $H(\theta)$ are given in the Supplemental Material~\cite{ThisSM}. Hence, at fixed $r=\phr_\dark/g$, interaction phase $\theta$, and readout noise $\measinp$, the number of measurements required for certification grows as $\nm_\star\propto g^{-1}$, even though the generated entanglement remains finite.

The number of measurements $\nm_\star$ does not by itself determine the time required for certification. In sequential release--recapture experiments, each measurement cycle involves preparation, a release interval of duration $\tint$, and readout. The total certification time is therefore $T_\star=\nm_\star(\tint+\trest)$, where $\trest$ accounts for preparation and readout. Combining Eq.~\eqref{eq:Nstar} with Eqs.~\eqref{eq:rr_sse_asymptotics} and \eqref{eq:rr_am_asymptotics}, we obtain
\begin{align}
    T_\star(\theta)
    \simeq \frac{\nsigma^2 \measinp^2 H(\theta)^2}{\epsilon^2[1-r F(\theta)]^2}
    \left(\frac{\theta}{\epsilon \trapfreq}+\trest\right).
    \label{eq:Tstar_RR_asymptotic}
\end{align}
At fixed $r$, $\measinp$, and $\theta$ with $rF(\theta)<1$, the number of measurements grows as $\epsilon^{-2}$, while the release duration increases as $\epsilon^{-1}$. When the release duration dominates the measurement cycle, the total certification time therefore scales as $T_\star\propto g^{-3/2}$. As the separability boundary is approached, $rF(\theta)\to1^{-}$, the certification time increases further due to the inverse-square dependence on the entanglement gap, in agreement with Eq.~\eqref{eq:Nstar}. However, varying the release duration affects not only the generated entanglement but also the statistical fluctuations and the time required per measurement cycle. The release duration that minimizes certification time therefore need not maximize the generated entanglement.

\begin{figure}
    \centering
    \includegraphics[width=\linewidth]{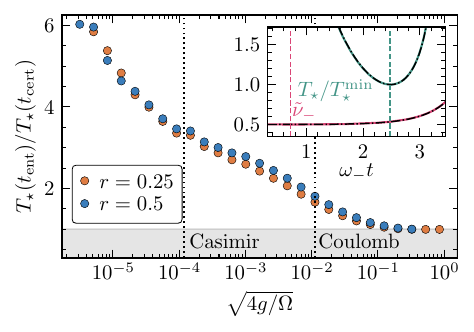}
    \caption{Ratio $T_\star(t_{\rm ent})/T_\star(t_{\rm cert})$ as a function of the weak-coupling parameter $\epsilon=\sqrt{4g/\trapfreq}$ for $r=0.25$ and $r=0.5$, with $\nbar=0.65$, $\measinp=1$, and $\trapfreq\trest=420\pi$. Both $t_{\rm ent}$ and $t_{\rm cert}$ are optimized over the interval $0<\freq_-t\leq2\pi$. Vertical dotted lines indicate the Casimir and Coulomb reference coupling strengths. Inset: entanglement witness and normalized certification time for $r=0.5$ and $\epsilon=10^{-5}$ with vertical dashed lines marking their respective minima. Black dash-dotted curves show the weak-coupling predictions, and the certification-time curve is normalized by its own minimum. The remaining parameters are taken from Table~\ref{tab:physical_params} of the End Matter.}
    \label{fig:time}
\end{figure}

To quantify the advantage of optimizing the release duration, we define $t_{\rm ent}$ and $t_{\rm cert}$ as the times minimizing $\sse(t)$ and $T_\star(t)$, respectively, within an experimentally accessible release window, which may be limited by the available free-fall time. As shown in the inset of Fig.~\ref{fig:time}, the certification time continues to decrease beyond the point of maximum entanglement, giving $t_{\rm cert}>t_{\rm ent}$. This occurs because the reduction in the fluctuation coefficient $\alpha$ compensates for both the smaller entanglement gap and the longer measurement cycle. A state with less entanglement can therefore be certified in less time. The main panel shows the ratio $T_\star(t_{\rm ent})/T_\star(t_{\rm cert})$, revealing a severalfold reduction in certification time in the weak-coupling regime for both diffusion ratios considered. This improvement requires only adjusting the release duration, while keeping the interaction strength, initial occupation, and measurement scheme fixed. The Coulomb and Casimir reference lines indicate representative interaction strengths. 

An alternative to release--recapture is to keep the particles trapped and continuously monitor their positions. The measurement record can be used to apply feedback that stabilizes the motion and enhances Einstein--Podolsky--Rosen correlations~\cite{Winkler2025}. The entanglement of the resulting stationary Gaussian state can then be certified from a sufficiently long measurement record~\cite{WisemanMilburn2009, Jacobs2014}. Although this avoids repeated preparation, measurement backaction acts throughout the dynamics and temporal correlations limit the rate at which independent information is acquired. In continuous monitoring, the unconditional covariance is $\cov_\uncond=\cov_\cond+\cov_\excess$, where $\cov_\cond$ is determined by the calibrated conditional dynamics and $\cov_\excess$ is the covariance of the fluctuating conditional mean, estimated from the measurement record. For a sufficiently long stationary record of duration $T$, the statistical uncertainty of the entanglement witness takes a form analogous to release--recapture~\cite{ThisSM}
\begin{align}
    \Std[\estimate{\tilde{\nu}}_-] \simeq \frac{\beta}{\sqrt{T/\corrtime}},
    \label{eq:cm_sse_fluct}
\end{align}
where $\beta^2=2\trc[\matG(\cov_\uncond)\cov_\excess\matG(\cov_\uncond)\cov_\excess]$ determines the fluctuation scale. The effective correlation time is defined as~\cite{ThisSM}
\begin{align}
    \corrtime = \frac{2}{\beta^2}\int_{-\infty}^{\infty}d\tau
    \trc[\matG(\cov_\uncond) \matR(\tau) \matG(\cov_\uncond) \matR^\transpose(\tau)],
\end{align}
where $\matR_{ij}(\tau)=\expect[\bar{\psvec}_{\cond,i}(t+\tau)\bar{\psvec}_{\cond,j}(t)]$ is the stationary autocorrelation matrix of the conditional mean $\bar{\psvec}_{\cond}$. Thus, $\corrtime$ accounts for temporal correlations, so that a record of duration $T$ provides the equivalent of $T/\corrtime$ independent measurements. Applying the same $\nsigma$-sigma certification criterion as for release--recapture gives
\begin{align}
    T_\star \simeq \left(\frac{\nsigma\beta}{1-\sse}\right)^2\corrtime.
\end{align}
The certification time therefore depends on both the statistical fluctuations and the rate of independent information acquisition. Hereafter, we adopt for concreteness the linear--quadratic--Gaussian (LQG) feedback strategy of Ref.~\cite{Winkler2025}, with fixed control parameters and a single force channel acting unequally on the two particles to enhance Einstein--Podolsky--Rosen correlations~\cite{ThisSM}.

\begin{figure}
    \centering
    \includegraphics[width=\linewidth]{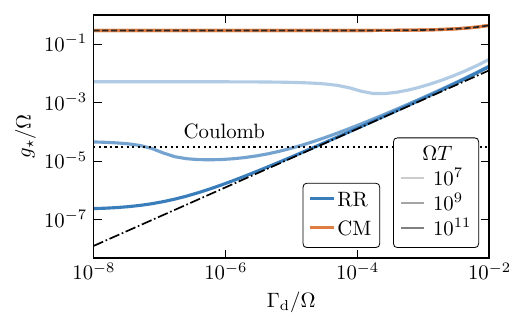}
    \caption{Minimum certifiable coupling $g_\star(T)$ as a function of the dark-diffusion rate for release--recapture (RR) at fixed release phase $\theta=\pi$ and continuous monitoring (CM) strategies, with $\nsigma=5$ and for different total experimental times $\trapfreq T=10^7,10^9,$ and $10^{11}$. The black dash-dotted line shows the weak-coupling entanglement-generation asymptote $\pi\phr_\dark/\sqrt{6}$ for release--recapture. The black dashed line shows the entanglement-generation boundary for continuous monitoring. Parameters are $\nbar=0.65$, $\trapfreq\trest=420\pi$, $\phr_\backaction/\trapfreq=0.05$, $\efficiency_\detection = 1$, zero intrinsic friction $\fric = 0$, and LQG control cost $\trapfreq q=0.1$. The actuation ratio for the LQG control is $5:1$. For release-recapture, the readout noise is $\measinp=\sqrt{1+\phr_\dark/\phr_\backaction}$. The dotted black line marks the Coulomb reference value evaluated using the parameters in Table~\ref{tab:physical_params} of the End Matter.}
    \label{fig:gmin}
\end{figure}

We now ask how weak an interaction can still produce certifiable entanglement within an experimental time budget $T$. For each protocol, keeping all other experimental parameters fixed, we define the minimum certifiable coupling as
\begin{align}
    g_\star(T)=\inf\left\{g>0:\sse(g)<1,\quad T_\star(g)\leq T\right\}.
    \label{eq:gmin}
\end{align}
As $T\to\infty$, this threshold approaches the minimum coupling required to generate entanglement, provided certification is possible in finite time. Figure~\ref{fig:gmin} reveals different limitations for the two protocols. For continuous monitoring, the certification threshold remains close to the entanglement generation boundary over the parameter range shown, indicating that the main limitation is generating an entangled unconditional steady state. Release--recapture enables certification at weaker couplings, although at low dark diffusion its threshold depends strongly on the available time budget. In this regime, further reducing dark diffusion provides limited improvement, as the time required for certification becomes the dominant constraint. At higher diffusion rates, the release--recapture curves approach the weak-coupling generation threshold $g_{\rm ent}\simeq\pi\phr_\dark/\sqrt{6}$ for a fixed release phase $\theta=\pi$~\cite{ThisSM}. This threshold marks the onset of entanglement, while finite-time certification requires stronger coupling to overcome statistical fluctuations. Optimizing the release duration can further reduce certification time, as shown in Fig.~\ref{fig:time}, but is not included in this comparison.

In summary, we have quantified how the measurement statistics of Gaussian states determine the time required to certify their entanglement. We derive an attainable lower bound on the number of measurements, which diverges as the inverse square of the distance to separability. The actual number also depends on the state covariance matrix and the available readout. Combining this statistical requirement with the time needed to acquire independent information allows us to compare release--recapture and continuous monitoring on equal footing. For the implementations considered, release--recapture enables certification at weaker interaction strengths, while optimizing the release duration can reduce certification time without maximizing entanglement. At low dark diffusion, the minimum certifiable coupling is limited mainly by the available experimental time budget, so further improvements in isolation provide little benefit. At higher decoherence rates, environmental noise becomes the dominant limitation, and reducing it extends the range of certifiable interactions. These results highlight the importance of accounting for certification resources when optimizing entanglement experiments, rather than maximizing entanglement generation alone.


\begin{acknowledgments}
We thank the Q-Xtreme synergy group for fruitful discussions. This research has been supported by the European Research Council (ERC) under the grant agreement No. [951234] (Q-Xtreme ERC-2020-SyG).
\end{acknowledgments}

\bibliography{bibliography}

\section*{End Matter}\label{EndMatter}

\setcounter{equation}{0}
\renewcommand{\theequation}{A\arabic{equation}}

\noindent
\emph{Appendix A: Universal lower bound on certification fluctuations}--- We derive the universal lower bound underlying Eq.~\eqref{eq:universal_std}. We assume a nondegenerate partially transposed symplectic spectrum, $\sse<\tilde{\nu}_{+}$, such that $\sse$ is differentiable.

Let $\Lambda=\operatorname{diag}(1,1,1,-1)$ implement partial transposition, and let $\matS$ symplectically diagonalize the partially transposed covariance matrix,
\begin{align}
    \matD
    &=\matS^{-1}\Lambda\cov\Lambda^\transpose
    \matS^{-\transpose} = \operatorname{diag}
    (\sse,\sse,\tilde{\nu}_{+},\tilde{\nu}_{+}).
    \label{eq:EM_williamson}
\end{align}
Let $\matP_{-}$ be the $2\times4$ matrix selecting the $\sse$ subspace, and define $\matW=\matP_{-}\matS^{-1}\Lambda$. First-order perturbation theory gives
\begin{align}
    d\sse &= \trc[\matW d\cov \matW^\transpose]/2 =\trc[\matG(\cov)d\cov],
\end{align}
from which we identify
\begin{align}
    \matG(\cov)= \matW^\transpose\matW/2.
\label{eq:EM_G}
\end{align}
Thus, $\matG(\cov)$ is positive semidefinite and has
rank two. It further satisfies
\begin{align}
    \matG(\cov)\cov\matG(\cov)
    =\frac{\sse}{2}\matG(\cov).
    \label{eq:EM_GVG}
\end{align}

We next establish a lower bound on $\trc[\matG(\cov)]$. Let $v_1$ and $v_2$ denote the first two rows of $\matS^{-1}$. Writing the symplectic form as $\matJ=\matJ_2\oplus\matJ_2$, with 
\begin{align}
    \matJ_2=\begin{pmatrix}0&1\\-1&0\end{pmatrix},
\end{align}
the symplectic relation for $\matS$ implies $|v_1\matJ v_2^\transpose|=1$. Using the arithmetic--geometric mean and Cauchy--Schwarz inequalities, we obtain
\begin{align}
    \trc[\matG(\cov)] &=\frac{\lVert v_1\rVert^2+\lVert v_2\rVert^2}{2} \geq\lVert v_1\rVert\lVert v_2\rVert \geq |v_1\matJ v_2^\transpose|=1.
    \label{eq:EM_trace_bound}
\end{align}

To obtain the bound quoted in the main text, let $g_1$ and $g_2$ be the two nonzero eigenvalues of $\matG(\cov)$. Since $g_1+g_2=\trc[\matG(\cov)]\geq1$, the fluctuation coefficient for isotropic measurement noise can be written as
\begin{align}
    \alpha^2 =\frac{1}{2}\left[ (\sse+2\measinp g_1)^2 +(\sse+2\measinp g_2)^2 \right].
    \label{eq:EM_alpha_eigenvalues}
\end{align}
For $\measinp\geq0$, this expression is minimized under the constraint $g_1+g_2\geq1$ at $g_1=g_2=1/2$, yielding
\begin{align}
    \alpha\geq\sse+\measinp.
    \label{eq:EM_alpha_bound}
\end{align}
Together with Eq.~\eqref{eq:Nstar}, this gives the lower bound on the number of measurements in Eq.~\eqref{eq:universal_std}. Further details of the gradient derivation and examples illustrating the dependence on the state covariance matrix are given in the Supplemental Material~\cite{ThisSM}.

\setcounter{equation}{0}
\renewcommand{\theequation}{B\arabic{equation}}

\begin{table}[!htb]
\caption{Reference parameters and coupling strengths for
Coulomb, Casimir--Polder, and gravitational interactions
between two levitated spheres.
\label{tab:physical_params}}
\begin{tabular}{ll}
\hline\hline
Parameter & Value \\
\hline
Radius $R$ & $100 \mathrm{nm}$ \\
Density $\rho$ & $2201 \mathrm{kg m^{-3}}$ \\
Mass $M_{\rm p}$ & $9.22\times10^{-18} \mathrm{kg}$ \\
Center separation $d$ & $1 \mu\mathrm{m}$ \\
Trap frequency $\trapfreq/(2\pi)$ & $100 \mathrm{kHz}$ \\
Zero-point displacement $\zppos$ & $3.02 \mathrm{pm}$ \\
Relative permittivity $\varepsilon_r$ & $2$ \\
Electric charges $(Q_\sysa,Q_\sysb)$ & $(+e,-e)$ \\
\hline
Interaction & $g/\trapfreq$ \\
\hline
Coulomb & $3.17\times10^{-5}$ \\
Casimir--Polder & $3.48\times10^{-9}$ \\
Gravitational & $7.79\times10^{-22}$ \\
\hline\hline
\end{tabular}
\end{table}

\noindent
\emph{Appendix B: Reference interaction strengths.---} To relate the dimensionless coupling $g/\trapfreq$ to physical interactions, we consider two identical levitated spheres with the parameters listed in Table~\ref{tab:physical_params}. For a central inverse-power-law potential $V_{\rm int}(r)=C_0/r^n$, with $n>0$, expansion to quadratic order in the transverse relative displacement gives
\begin{align}
    \frac{g}{\trapfreq}
    =-n\frac{V_{\rm int}(d)}{\hbar\trapfreq}
    \left(\frac{\zppos}{d}\right)^2,
    \label{eq:EM_physical_coupling}
\end{align}
where $d$ is the equilibrium center-to-center separation and $\zppos$ is the zero-point displacement.

The Coulomb and Casimir--Polder coupling strengths are indicated in Fig.~\ref{fig:time}, while the Coulomb value is also shown in Fig.~\ref{fig:gmin} for comparison with the minimum certifiable coupling. These values serve as physical reference scales and are not imposed when varying $g/\trapfreq$ in the certification analysis. Explicit expressions for the three interactions are given in the Supplemental Material~\cite{ThisSM}.

\end{document}


\title{Supplemental Material:\\
Certification Time for Weak-Force-Mediated Entanglement between Levitated Particles}

\author{Yao-Tong Chen \orcid{0000-0003-3687-6541}}
\affiliation{ICFO – Institut de Ciencies Fotoniques, The Barcelona Institute of Science and Technology, Castelldefels, Barcelona 08860, Spain}

\author{Salambô Dago \orcid{0000-0002-1964-2375}}
\affiliation{University of Vienna, Faculty of Physics, Vienna Center for Quantum Science and Technology (VCQ), Boltzmanngasse 5, Vienna, 1090, Austria}

\author{Oriol Romero-Isart \orcid{0000-0003-4006-3391}}
\affiliation{ICFO – Institut de Ciencies Fotoniques, The Barcelona Institute of Science and Technology, Castelldefels, Barcelona 08860, Spain}
\affiliation{ICREA – Institucio Catalana de Recerca i Estudis Avançats, Barcelona 08010, Spain}

\author{Andreu Riera-Campeny \orcid{0000-0003-3260-993X}}
\affiliation{ICFO – Institut de Ciencies Fotoniques, The Barcelona Institute of Science and Technology, Castelldefels, Barcelona 08860, Spain}

\date{\today}

\maketitle

\setcounter{figure}{0}
\setcounter{equation}{0}
\renewcommand{\thefigure}{S\arabic{figure}}
\renewcommand{\theequation}{S\arabic{equation}}
\renewcommand{\bibnumfmt}[1]{[S#1]}
\renewcommand{\citenumfont}[1]{S#1}
\setcounter{section}{0}
\renewcommand\thesection{S\arabic{section}}

In this Supplemental Material, we derive several mathematical results used in the main text.

\section{Mechanical Interaction Model}
\label{sec:physical_framework}

We consider two particles interacting through a central inverse-power-law potential
\begin{align}
    V_{\rm int}(\rvec)=\frac{C_0}{r^n}=\frac{C_0}{(r_x^2+r_y^2+r_z^2)^{n/2}},
\end{align}
where $\rvec=(r_x,r_y,r_z)$ is the relative position vector, $r = (r_x^2+r_y^2+r_z^2)^{1/2}$, $n>0$, and $C_0$ sets the interaction strength. This form includes gravitational, Coulomb, and Casimir--Polder interactions as special cases. We focus on the regime where the motional amplitudes remain small compared to the equilibrium separation $d$, allowing the interaction to be expanded around equilibrium. Writing
\begin{align}
    \rvec = \rvec_d + \delta \rvec,
\end{align}
with $\rvec_d=(d,0,0)$ and $\delta \rvec=(\delta r_x,\delta r_y,\delta r_z)$, the interaction potential can be expanded to second order as
\begin{align}
    V_{\rm int}(\rvec)\approx\frac{C_0}{d^n}\left[1-\frac{n\delta r_x}{d}+\frac{(n^2+n)\delta r_x^2-n(\delta r_y^2+\delta r_z^2)}{2d^2}\right].
\end{align}

To this order, the dynamics along different spatial directions decouple. Since optical readout is typically performed along a direction transverse to the interparticle axis, we focus on the corresponding transverse mode. The interaction then reduces to
\begin{align}
\label{eq:linearized_interaction}
    \hat{V} = -\frac{n C_0 \zppos^2}{2 d^{n+2}} (\posop_\sysa-\posop_\sysb)^2 \equiv \frac{\hbar g}{2} (\posop_\sysa-\posop_\sysb)^2,
\end{align}
where $\zppos$ is the zero-point motion, and $g$ defines the effective coupling rate. In general, we see 
\begin{align}
    \frac{g}{\trapfreq} = -n \frac{\pot_{\rm int}(d)}{\hbar\trapfreq} \left(\frac{\zppos}{d}\right)^2.
\end{align}

More precisely, considering the gravitational, Coulomb, and Casimir-Polder interactions between two levitated nanoparticles at a distance $d$, we find~\cite{Emig2007, Weiss2021} 
\begin{align}
    &\frac{g_{\rm G}}{\trapfreq} = \frac{G \mass^2}{d \hbar\trapfreq} \left(\frac{\zppos}{d}\right)^2,\\
    &\frac{g_{\rm C}}{\trapfreq} = -\frac{1}{4\pi\varepsilon_0}\frac{Q_{\rm A} Q_{\rm B}}{\hbar \trapfreq d} \left(\frac{\zppos}{d}\right)^2,\\
    &\frac{g_{\rm Ca}}{\trapfreq} = \frac{7 \times 23}{4\pi} \frac{c}{\trapfreq} \left(\frac{\varepsilon_{r}-1}{\varepsilon_{r}+2}\right)^2 \frac{R^6}{d^7} \left(\frac{\zppos}{d}\right)^2,
\end{align}
respectively, where $G$ is Newton's gravitational constant, $\varepsilon_0$ is the vacuum permittivity, $\varepsilon_r$ is the relative permittivity, $Q_s$ is the charge of system $s$, and $R$ is the particle radius.

\section{Asymptotic fluctuations of the witness estimator}
\label{app:universal_rate}
We derive identities relating the gradient matrix $\matG(\cov)$ to the covariance matrix $\cov$ and use them to characterize the asymptotic fluctuations of the witness estimator. Throughout this section, we assume that the symplectic eigenvalues of the partially transposed covariance matrix are distinct, $\sse < \tilde{\nu}_+$, so that $\sse$ is differentiable.

\subsection{The gradient matrix $\matG(\cov)$}

For any two-mode Gaussian state, the smallest symplectic eigenvalue of the partially transposed covariance matrix, $\sse$, is fully determined by two symplectic invariants, $\tilde{\Delta}\equiv \det(\matA)+\det(\matB)-2\det(\matC)$ and $\det(\cov)$. Differentiating the expression for $\sse$ in the main text gives the differential relation
\begin{align}
d[\sse] = \frac{-\sse^{2} d[\tilde{\Delta}] + d[\det(\cov)]}{2\sse \sqrt{\tilde{\Delta}^{2}-4\det(\cov)}},
\label{eq:dnu_invariants}
\end{align}
which makes explicit that infinitesimal fluctuations of $\sse$ depend solely on fluctuations of these two invariants.
By definition, the gradient satisfies $d[\sse] = \trc[\matG(\cov) d\cov]$. This relation allows us to identify $\matG(\cov)$ by expressing the differentials of the symplectic invariants in terms of $d\cov$. Using the Jacobi identity $d[\det(\matM)] = \det(\matM) \trc[\matM^{-1} d\matM]$, we obtain
\begin{align}
&d[\tilde{\Delta}] = \trc[\matL d\cov], \label{eq:tilde_Delta_differential}\\
&d[\det(\cov)] = \det(\cov) \trc[\cov^{-1} d\cov],\label{eq:tilde_detV_differential}
\end{align}
where we have introduced
\begin{align}
\matL = \begin{pmatrix}
\det(\matA) \matA^{-1} & -\det(\matC) \matC^{-1 \transpose}\\
-\det(\matC) \matC^{-1} & \det(\matB) \matB^{-1}
\end{pmatrix}.
\end{align}
If $\matC$ is singular, the products $\det(\matC)\matC^{-1}$ and $\det(\matC)\matC^{-\transpose}$ are understood as the adjugate of $\matC$ and its transpose, respectively. Combining these expressions we obtain a closed expression for the gradient $\matG(\cov)$ in the form
\begin{align}
\matG(\cov) = \frac{-\sse^{2} \matL + \det(\cov)\cov^{-1}}
{2\sse\sqrt{\tilde{\Delta}^{2}-4\det(\cov)}}.
\label{eq:grad_cov}
\end{align}

From its definition, one can check that $\sse$ is a homogeneous function of degree one. Therefore, Euler's theorem gives
\begin{align}
\label{eq:homogeneous_function}
    \sse = f(\cov)=\sum_{ij}\matG_{ij}(\cov)\cov_{ij}=\trc[\matG(\cov)\cov].
\end{align}
As a sanity check, this identity can be recovered explicitly using Eq.~\eqref{eq:grad_cov}.

\subsection{Symplectic algebraic representation of $\matG(\cov)$}

Let $\bbLambda = {\rm diag}(1,1,1,-1)$ be the matrix that implements the partial transposition. Let $\matS$ be the symplectic matrix that provides the symplectic diagonalization of $\bbLambda \cov \bbLambda^\transpose$, through
\begin{align}
\label{eq:symplectic_diagonalization}
    \matD = \matS^{-1} \bbLambda \cov \bbLambda^\transpose \matS^{-\transpose},
\end{align}
with $\matD = {\rm diag}(\sse,\sse,\tilde{\nu}_+,\tilde{\nu}_+)$. Let also $\matP_-$ be the $2\times4$ matrix that selects the subspace of the eigenvalue $\sse$. It follows that the corresponding sub-block is
\begin{align}
\label{eq:d_minus}
    \matD_- = \sse \matI_2 = \matW \cov \matW^\transpose,
\end{align}
where $\matW = \matP_- \matS^{-1} \bbLambda$ and $\matD_- \equiv \matP_-\matD\matP_-^\transpose$. 

To determine the change in $\sse$ induced by $d\cov$, we work in the unperturbed symplectic basis, keeping $\matS$ fixed. The perturbation of the first diagonal block is then $d\matD_-=\matW d\cov\matW^\transpose$. Since $\sse<\tilde{\nu}_+$, coupling between the two symplectic modes contributes only at second order. To first order, the perturbed symplectic eigenvalue is therefore
\begin{align}
    \sqrt{\det(\sse \matI_2 + d\matD_-)} \simeq \sse + \trc[d\matD_-]/2 \equiv \sse + d[\sse]. 
\end{align}
Expressing $\matD_-$ as a function of the covariance matrix through Eq.~\eqref{eq:d_minus}, we find
\begin{align}
    d[\sse] = \trc[\matG(\cov) d\cov] = \trc[\matW d\cov \matW^\transpose]/2, 
\end{align}
and therefore we identify $\matG(\cov) = \matW^\transpose \matW/2$. Note that this expression makes explicit that $\matG(\cov)$ is a positive-semidefinite rank-2 matrix. The following crucial property follows
\begin{align}
\label{eq:crucial_property}
    &\matG(\cov) \cov \matG(\cov) = \sse \matG(\cov)/2. 
\end{align}

\subsection{A lower bound on the trace of $\matG(\cov)$}

We now show that, in addition to being positive semidefinite, $\matG(\cov)$ has a trace bounded from below by 1. Indeed, let $\vvec_1$ and $\vvec_2$ be the first two rows of $\matS^{-1}$. First, we note that 
\begin{align}
    \trc[\matG(\cov)] = \frac{1}{2} \trc[\matP_-\matS^{-1} (\matP_-\matS^{-1})^\transpose] = \frac{\lVert\vvec_1\lVert^2 +  \lVert\vvec_2\lVert^2}{2},
\end{align}
where $\bbLambda^2 = \matI$ has dropped.  The symplectic relation for $\matS$ implies that $|\vvec_1 \cdot (\matJ \vvec_2)| = 1$, where $\matJ$ denotes the standard symplectic form. Then from the Cauchy-Schwarz inequality it follows that
\begin{align}
    1 = |\vvec_1 \matJ \vvec_2^\transpose|\leq
\lVert\vvec_1\rVert\lVert\matJ\vvec_2^\transpose\rVert= \lVert\vvec_1\lVert\lVert\vvec_2\lVert.
\end{align}
Finally, the arithmetic-mean--geometric-mean inequality implies that 
\begin{align}
    \trc[\matG(\cov)] = \frac{\lVert\vvec_1\lVert^2 +  \lVert\vvec_2\lVert^2}{2} \geq \lVert\vvec_1\lVert \lVert\vvec_2\lVert \geq 1.
\end{align}
This shows that $\matG(\cov)$ cannot be made arbitrarily small, and ultimately will bound the fluctuations of the asymptotic estimate of $\sse$.

\subsection{Asymptotic fluctuations of the estimate}

In the main text we have established that, in the asymptotic limit $\nm \gg 1$, the fluctuations of the estimator $\estimate{\tilde{\nu}}_-$ are given by
\begin{align}
\label{eq:generic_var_sse}
    \nm \Var[\estimate{\tilde{\nu}}_-] \sim 2 \trc[\matG(\cov)\bbSigma\matG(\cov)\bbSigma],
\end{align}
where $\bbSigma = \cov + \cov_\meas$ is the measured covariance matrix. For heterodyne or retrodiction measurement schemes, the added measurement noise is $\cov_\meas = \measinp \matI$, for some measurement noise amplitude $\measinp$. Using Eq.~\eqref{eq:crucial_property} it follows that
\begin{align}
\label{eq:asymptotic_sse_grad}
    \nm \Var[\estimate{\tilde{\nu}}_-] \sim \sse^2 + 2 \measinp \sse \trc[\matG(\cov)] + 2 \measinp^2 \trc[\matG(\cov)^2].
\end{align}
In Eq.~\eqref{eq:asymptotic_sse_grad}, the first term depends solely on the entanglement, while the second and third depend on the amount of measurement noise and the state $\cov$.

We would like to obtain a universal lower bound that depends only on the amount of entanglement through $\sse$ but not on the particular state $\cov$. Let $g_1, g_2>0$ be the non-zero eigenvalues of $\matG(\cov)$, which satisfy $g_1 + g_2 \geq 1$. It follows that 
\begin{align}
\label{eq:asymptotic_sse_evals}
    \nm \Var[\estimate{\tilde{\nu}}_-] \sim \frac{(\sse + 2 \measinp g_1)^2 + (\sse + 2 \measinp g_2)^2}{2},
\end{align}
which is minimized for $g_1 = g_2 = 1/2$. Hence, we find that 
\begin{align}
\label{eq:asymptotic_bound}
    \sqrt{\nm} \Std[\estimate{\tilde{\nu}}_-] \gtrsim \sse + \measinp.
\end{align}
It is easy to check that this bound is saturated by symmetric thermal two-mode squeezed states.

\section{A toy model: Fixed entanglement with increasing fluctuations}
\label{sec:sm_local_squeezing}

We consider a family of Gaussian states related by local squeezing operations that preserve entanglement while making the asymptotic uncertainty of its covariance-based estimate arbitrarily large. The family is parametrized by the local squeezing $u$ applied to some given initial state. As in the main text, the measurement outcomes are independent Gaussian samples with covariance $\bbSigma(u)=\cov(u)+\cov_\meas$, where the calibrated measurement noise $\cov_\meas=\measinp\matI$ is held fixed.

The family is defined operationally, starting from a two-mode squeezed vacuum state with fixed squeezing parameter $u_0>0$. Defining $c=\cosh(u_0)$, $s=\sinh(u_0)$, and $\matZ=\operatorname{diag}(1,-1)$, its covariance matrix is
\begin{align}
    \cov(0)=\begin{pmatrix}
        c\matI_2 & s\matZ\\
        s\matZ & c\matI_2
    \end{pmatrix}.
\end{align}
Its smallest partially transposed symplectic eigenvalue is $\sse=c-s=e^{-u_0}$, and the corresponding gradient matrix is
\begin{align}
    \matG(0)=\frac{1}{4}\begin{pmatrix}
        \matI_2 & -\matZ\\
        -\matZ & \matI_2
    \end{pmatrix}.
\end{align}

We then apply opposite local squeezings parametrized by $u\in\mathbb{R}$, described by the symplectic matrix
\begin{align}
    \matS(u)=\operatorname{diag}(e^u,e^{-u},e^{-u},e^u).
\end{align}
The resulting family has covariance $\cov(u)=\matS(u)\cov(0)\matS(u)^\transpose$. Since these transformations are local, they leave $\sse$ unchanged. Applying the chain rule to $d\sse=\trc[\matG\,d\cov]$, the gradient transforms as $\matG(u)=\matS(u)^{-\transpose}\matG(0)\matS(u)^{-1}$, giving
\begin{align}
    \matG(u)=\frac{1}{4}\begin{pmatrix}
        e^{-2u} & 0 & -1 & 0\\
        0 & e^{2u} & 0 & 1\\
        -1 & 0 & e^{2u} & 0\\
        0 & 1 & 0 & e^{-2u}
    \end{pmatrix},
\end{align}
where $\matS(u)^{-\transpose}\equiv[\matS(u)^{-1}]^\transpose$.

To evaluate the asymptotic fluctuation coefficient $\alpha$, we use
\begin{align}
    \trc[\matG(u)]&=\cosh(2u),\\
    \trc[\matG(u)^2]&=\cosh^2(2u)/2.
\end{align}
Substituting these expressions into the general fluctuation formula gives
\begin{align}
    \alpha(u)^2
    &=\sse^2+2\measinp\sse\trc[\matG(u)]
    +2\measinp^2\trc[\matG(u)^2]\nonumber\\
    &=[\sse+\measinp\cosh(2u)]^2.
\end{align}
The number of measurements required for certification at $\nsigma$ standard deviations is therefore
\begin{align}
    \nm_\star(u)\simeq
    \frac{\nsigma^2[\sse+\measinp\cosh(2u)]^2}
    {(1-\sse)^2}.
\end{align}
For fixed $u_0>0$ and $\measinp\geq1$, this requirement grows without bound as $|u|$ increases, although the entanglement remains unchanged. Local squeezing changes how entanglement is encoded in the covariance matrix, increasing the sensitivity of its estimation to the fixed measurement noise. This conclusion applies to the covariance-based estimator and readout considered here. Reversing the local squeezing before measurement restores the original certification requirements.

\section{Entanglement dynamics in the release--recapture model}

We first analyze the coherent dynamics during the release stage, assuming that the system is initially in the ground state, $\cov(0)=\matI$. An analytical description is obtained by introducing the collective quadratures $\posop_\pm=(\posop_\sysa\pm\posop_\sysb)/\sqrt{2}$ and $\momop_\pm=(\momop_\sysa\pm\momop_\sysb)/\sqrt{2}$. This symplectic transformation decouples the dynamics into two independent modes. In these variables, the Hamiltonian reads
\begin{align}
\Hop = \frac{\hbar\trapfreq}{4} \momop_+^2 + \frac{\hbar\trapfreq}{4} \left(\momop_-^2 + \frac{4g}{\trapfreq} \posop_-^2\right),
\end{align}
describing a free common mode and a harmonic differential mode with frequency $\freq_-=\sqrt{4g\trapfreq}$. The dimensionless ratio $4g/\trapfreq$ distinguishes the weak-coupling ($4g<\trapfreq$) and strong-coupling ($4g>\trapfreq$) regimes.

\subsection{Exact coherent dynamics}

The common mode evolves freely, producing a phase-space shear. Its second moments are
\begin{align}
&\av{\posop_+^2}(t) = 1 + (\trapfreq t)^2,\\
&\av{\momop_+^2}(t) = 1,\\
&\av{\{\posop_+,\momop_+\}}(t)/2 = \trapfreq t.
\end{align}
Consequently, its phase-space distribution is sheared into an increasingly elongated ellipse, whose squeezed axis asymptotically approaches the $\momop_+$ direction. The differential mode instead undergoes an elliptical phase-space rotation, with second moments
\begin{align}
&\av{\posop_-^2}(t) = \cos^2(\freq_- t) + \frac{\trapfreq}{4g}\sin^2(\freq_- t),\\
&\av{\momop_-^2}(t) = \cos^2(\freq_- t) + \frac{4g}{\trapfreq}\sin^2(\freq_- t),\\
&\av{\{\posop_-,\momop_-\}}(t)/2 = \left(1 - \frac{4g}{\trapfreq}\right)\frac{\trapfreq}{2\freq_-}\sin(2\freq_- t),
\end{align}
with $\freq_- = \sqrt{4g\trapfreq}$ the differential mode frequency.

These collective-mode dynamics explain the different timing of entanglement generation in the weak- and strong-coupling regimes. The common mode develops squeezing along a quadrature approaching $\momop_+$, while EPR-type correlations require sufficiently small fluctuations in the complementary differential quadrature. In the weak-coupling regime $4g<\trapfreq$, relevant here, this quadrature is close to $\posop_-$, and entanglement is therefore approximately maximized when $\Var(\posop_-)$ is minimal. This occurs at integer multiples of the half-period $\tpi=\pi/\freq_-$. 


\subsection{Open dynamics in the weak-coupling limit}

We now consider the general open system dynamics, with an initial thermal state with mean phonon occupation $\nbar$. We are interested in the entanglement dynamics mediated by weak forces and, consequently, we consider the weak-coupling limit where $\epsilon=\sqrt{4g/\trapfreq}$ is our small parameter. Simultaneously, we keep $r=\phr_\dark/g >0$ fixed and of order unity. With this definition we can replace $4\phr_\dark/\trapfreq \mapsto r\epsilon^2$. This scaling retains the competition between entanglement generation and dark momentum diffusion. We also hold the initial thermal occupation $\nbar$ and the interaction phase $\theta=\freq_-t$ fixed, thereby resolving the release dynamics on the interaction timescale. Note that this expansion is not uniform as $r\to0$ or $\theta \to 0$. Defining $\mu=2\nbar+1$, the common-mode second moments become
\begin{align}
&\av{\posop_+^2} = \frac{\mu\theta^2}{\epsilon^2}+\frac{r\theta^3}{3\epsilon}+\mu,\\
&\av{\momop_+^2} = \mu+r\theta\epsilon,\\
&\frac{\av{\{\posop_+,\momop_+\}}}{2} = \frac{\mu\theta}{\epsilon}+\frac{r\theta^2}{2},
\end{align}
and the differential-mode second moments are
\begin{align}
&\av{\posop_-^2} = \frac{\mu\sin^2\theta}{\epsilon^2}+\frac{r}{2\epsilon}\left(\theta-\frac{\sin2\theta}{2}\right)+\mu\cos^2\theta,\\
&\av{\momop_-^2} = \mu\cos^2\theta+\frac{r\epsilon}{2}\left(\theta+\frac{\sin2\theta}{2}\right)+\mu\epsilon^2\sin^2\theta,\\
&\frac{\av{\{\posop_-,\momop_-\}}}{2} = \frac{\mu\sin2\theta}{2\epsilon}+\frac{r\sin^2\theta}{2}-\frac{\mu\epsilon\sin2\theta}{2}.
\end{align}
We note that the leading terms are those corresponding to position observables, going as $O(\epsilon^{-2})$. These expressions are exact and organize the second moments by their powers of $\epsilon$ at fixed $\mu$, $r$, and $\theta$. Terms that are sub-leading in individual matrix elements must be retained when evaluating the symplectic invariants, since their leading contributions can cancel.

To evaluate the entanglement dynamics and certification cost, we express $\sse$ and $\matG(\cov)$ in the collective-mode basis. Let $\cov_+$ and $\cov_-$ denote the two-by-two covariance matrices of the common and differential modes, respectively. In this basis, the symplectic invariants entering $\sse$ are $\tilde{\Delta} = \trc[{\rm adj}(\cov_+) \cov_-]$ and $\det(\cov) = \det(\cov_+) \det(\cov_-)$. The gradient matrix is block diagonal, with blocks
\begin{align}
    \matG_\pm = \frac{\det(\cov_\mp){\rm adj}(\cov_\pm)-\sse^2 {\rm adj}(\cov_\mp)}{2\sse \zeta},
\end{align}
where $\zeta = (\tilde{\Delta}^2-4\det(\cov_+) \det(\cov_-))^{1/2}$. These expressions apply when the partially transposed symplectic eigenvalues are nondegenerate, so that $\zeta>0$.

We now derive the leading contributions to $\sse$ and the asymptotic fluctuation coefficient $\alpha$ as $\epsilon\to0$, keeping $\mu$, $r>0$, and $\theta>0$ fixed. The determinants satisfy
\begin{align}
    &\det(\cov_+) \sim \frac{\mu r \theta^3}{3\epsilon},\\
    &\det(\cov_-) \sim \frac{\mu r}{\epsilon} \frac{\theta - \sin(\theta) \cos(\theta)}{2},\\
    &\det(\cov) \sim \frac{\mu^2 r^2 \theta^3}{3\epsilon^2}\frac{\theta - \sin(\theta) \cos(\theta)}{2}.\label{eq:asymptotic_det}
\end{align}
The leading contribution to the other symplectic invariant is
\begin{align}
\label{eq:asymptotic_delta}
    \tilde{\Delta} \sim \frac{\mu^2}{\epsilon^2}[\sin\theta - \theta \cos(\theta)]^2.
\end{align}
This expansion requires $\sin(\theta)-\theta\cos(\theta)$ to remain nonzero as $\epsilon\to0$. Comparing Eqs.~\eqref{eq:asymptotic_det} and \eqref{eq:asymptotic_delta}, we find $\det(\cov)/\tilde{\Delta}^2=O(\epsilon^2)$ and hence $\zeta\sim\tilde{\Delta}$. The resulting leading expression for the witness is
\begin{align}
    \sse \sim r \frac{\sqrt{\theta^3 [\theta - \sin(\theta) \cos(\theta)]}}{\sqrt{6}|\theta \cos(\theta)-\sin\theta|} \equiv r F(\theta) .
\end{align}
Thus the leading witness is independent of $\epsilon$ and the initial thermal occupation. The expansion is valid away from the zeros of $\sin(\theta)-\theta\cos(\theta)$. In particular, it applies in a fixed neighborhood of $\theta=\pi$.

We next evaluate the absolute fluctuation coefficient $\alpha$, defined by $\Std[\estimate{\tilde{\nu}}_-]\simeq\alpha/\sqrt{\nm}$. For isotropic measurement noise,
\begin{align}
    \alpha^2 = \sse^2 + 2\measinp\sse \trc[\matG_+ + \matG_-] + 2\measinp^2\trc[\matG_+^2 + \matG_-^2].
\end{align}
Power counting shows that the diagonal momentum entries are the only elements of $\matG_\pm$ that scale as $O(\epsilon^{-1})$. Their leading contributions arise from $\det(\cov_\mp){\rm adj}(\cov_\pm)$. It therefore suffices to evaluate these two entries to obtain the leading fluctuation coefficient:
\begin{align}
    &\matG_{+,\mom\mom} \sim \frac{1}{\epsilon} \frac{3F(\theta)}{2\theta} \equiv \frac{\kappa_+(\theta)}{\epsilon},\\
    &\matG_{-,\mom\mom} \sim \frac{1}{\epsilon} \frac{\theta^3 \sin(\theta)^2}{6F(\theta)[\sin(\theta) - \theta\cos(\theta)]^2}\equiv \frac{\kappa_-(\theta)}{\epsilon},
\end{align}
where $\kappa_\pm(\theta)$ depend only on the interaction phase. For fixed $\measinp>0$, the term quadratic in the measurement noise dominates, giving
\begin{align}
    \alpha \sim \measinp H(\theta)/\epsilon,
\end{align}
with $H(\theta)^2 \equiv 2(\kappa_+(\theta)^2 + \kappa_-(\theta)^2)$. The leading witness therefore depends on $r$ but not on $\epsilon$, whereas the statistical uncertainty depends on $\epsilon$ but not on $r$. At $\theta=\pi$, the phase functions give $F(\pi)=\pi/\sqrt{6}$ and $H(\pi)=\sqrt{3}/2$, recovering the expressions quoted in the main text. 

\begin{figure}
    \centering
    \includegraphics[width=\linewidth]{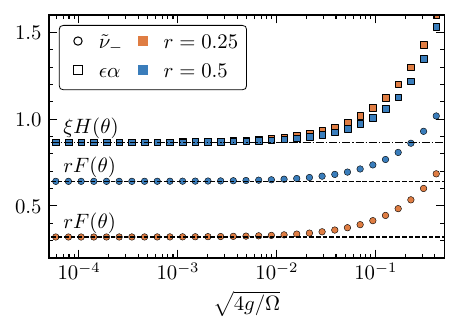}
    \caption{Finite entanglement with increasing readout sensitivity. Full covariance calculations of $\sse$ (circles) and $\epsilon\alpha$ (squares) at $\theta=\pi$ for $r=0.25$ (orange) and $r=0.5$ (blue), with $\nbar=0.65$ and $\measinp=1$. Dashed and dash-dotted lines show $rF(\pi)$ and $\measinp H(\pi)$, respectively. The plateau of $\epsilon\alpha$ implies $\alpha\propto g^{-1/2}$ at fixed $\trapfreq$, while $\sse$ approaches a finite value. Here the dark-diffusion rate $\phr_\dark$ scales with $g$ such that $r=\phr_\dark/g$ remains fixed.}
    \label{fig:asymptotics}
\end{figure}

Figure~\ref{fig:asymptotics} shows that the full dynamics approach these predictions: $\sse$ tends to a finite value, while $\epsilon\alpha$ approaches a nonzero plateau. For $rF(\theta)<1$, improved isolation and longer release therefore preserve finite entanglement as the force weakens, but its certification requires increasingly many measurements, $\nm_\star\propto\epsilon^{-2}$ at fixed readout noise. This statistical penalty is additional to the longer duration of each realization. Both leading expressions are independent of the fixed initial occupation. The expansion applies at fixed nonzero $r$ and $\sin\theta-\theta\cos\theta\neq0$ and is not uniform as either quantity vanishes.

\subsection{Measurement Model: Retrodiction}
\label{subsec:tms_retrodiction}

To certify the generated entanglement, we consider a retrodiction protocol~\cite{Zhang2017, Lammers2024}, particularly well suited to free-space levitated particles and recently demonstrated in single-particle settings~\cite{Rossi2025, Tomassi2026, Seta2026}. At the end of the release stage, both particles are recaptured in their optical tweezers. Assuming the interaction becomes negligible after recapture, each particle evolves independently under harmonic confinement at frequency $\trapfreq$. In this regime, the retrodiction problem admits an analytical solution: a sufficiently long continuous measurement record effectively implements a Gaussian measurement at time $\tint$, providing simultaneous information about the position and momentum quadratures of both particles.

Repeating the protocol $\nm$ times yields independent retrodicted phase-space outcomes
\begin{align}
    \psvec_n=(\pos_\sysa,\mom_\sysa,\pos_\sysb,\mom_\sysb),
\end{align}
for $n=1,\ldots,\nm$. For the zero-mean states considered here, the corresponding sample covariance is
\begin{align}
    \estimate{\bbSigma} = \frac{1}{\nm} \sum_{n=1}^{\nm} \psvec_n\psvec_n^\transpose = \estimate{\cov}(\tint)+\cov_\meas,
\end{align}
where $\cov_\meas$ denotes the additional covariance introduced by the retrodiction procedure. This contribution depends on the measurement $\phr_\meas$ and diffusion rates $\phr_\dark, \phr_\backaction$~\cite{Lammers2024}; in the ideal measurement limit, one has $\cov_\meas\to\matI$, corresponding to a projection onto coherent states.

Assuming $\cov_\meas$ is perfectly calibrated, the covariance matrix of the generated state can be estimated as
\begin{align}
    \check{\cov}(\tint) = \frac{1}{\nm} \sum_{n=1}^{\nm} \psvec_n\psvec_n^\transpose -\cov_\meas.
\end{align}
The certification procedure therefore relies on a calibrated measurement model and is not device independent, since calibration uncertainties propagate directly into the inferred covariance matrix.

\subsection{Estimates of timescales for the release--recapture protocol}
\label{app:timescales_rr}

In the release--recapture protocol, the duration of a single experimental realization $\runtime = \tint + \trest$ has a contribution coming from the interaction time $\tint$, but also a contribution $\trest = \tcool + \tmeas$, where $\tcool$ denotes the time required to cool the initial mechanical state and $\tmeas$ the time required to infer the final quadratures after recapture.

We estimate the cooling time by assuming that, after recapture or reloading, the particle can be cooled close to the critically damped regime. In this regime, the relaxation time is set by the mechanical period, up to a numerical factor depending on the cooling protocol and target occupation. As a conservative estimate, we take
\begin{align}
    \tcool=10\times \frac{2\pi}{\trapfreq}.
\end{align}
The factor of $10$ should be understood as a safety margin ensuring several relaxation times before the next release. 

The measurement time is set by the rate at which information about the mechanical quadratures is acquired during the retrodiction stage. Denoting the measurement rate by $\Gamma_\meas$, we similarly take
\begin{align}
    \tmeas = 10 \times \frac{2\pi}{\Gamma_\meas}.
\end{align}
Combining these estimates gives
\begin{align}
    \trest = 10 \times \left(\frac{2\pi}{\trapfreq}+\frac{2\pi}{\Gamma_\meas}\right).
\end{align}

This estimate is not intended as a fundamental quantity, but as a conservative parametrization of the experimental overhead associated with cooling and readout. It allows the sample complexity $\nm_\star$ to be translated into a total certification time through
\begin{align}
    T_\star = \nm_\star\left(\tint+\trest\right).
\end{align}

\section{Continuously monitored protocol}
\label{app:continuous_monitoring}

We consider two identical nanoparticles confined in independent optical tweezers and continuously monitored through the light scattered from the trapping beams. The measurement record provides information about the particle positions while inducing measurement backaction. Together with the local trapping potentials, the interaction in Eq.~\eqref{eq:linearized_interaction} yields the Hamiltonian
\begin{align}
\label{eq:cm_hamiltonian}
\Hop = \frac{\hbar\trapfreq}{4}(\momop_+^2+\posop_+^2)
+\frac{\hbar\trapfreq}{4}
\left(\momop_-^2+\frac{\trapfreq+4g}{\trapfreq}\posop_-^2\right),
\end{align}
where $\posop_\pm$ and $\momop_\pm$ denote the collective mechanical quadratures. In addition to the coherent dynamics generated by Eq.~\eqref{eq:cm_hamiltonian}, the particles are subject to measurement backaction, environmental decoherence, and feedback. Their interplay gives rise to a nonequilibrium Gaussian steady state whose entanglement can be inferred from the continuous measurement record.

\subsection{Conditional dynamics}
\label{app:conditional_dynamics}

We denote the measurement record acquired during $t\in[0,T]$ by $d\yvec(t)$. Conditioned on this record, the system is described by the state $\state_\cond$, which incorporates the information acquired through the measurement as well as its associated backaction~\cite{WisemanMilburn2009,Jacobs2014,Lammers2024}. Discarding the measurement record instead gives the unconditional state
\begin{align}
\state_\uncond=\av{\state_\cond}_{\rm tr},
\end{align}
where $\av{\cdot}_{\rm tr}$ denotes the ensemble average over measurement trajectories.

For the quadratic Hamiltonian in Eq.~\eqref{eq:cm_hamiltonian}, linear measurements, and linear feedback, the conditional dynamics preserve Gaussianity. The conditional state is therefore fully characterized by the mean $\bar{\xvec}_\cond=\av{\hat{\xvec}}_\cond$ and covariance matrix
\begin{align}
\cov_{\cond,ij}
=\frac{1}{2}
\av{
\left\{
\hat{\xvec}_i-\bar{\xvec}_{\cond,i},
\hat{\xvec}_j-\bar{\xvec}_{\cond,j}
\right\}
}_\cond .
\end{align}
Their evolution is governed by the Kalman-filter equations~\cite{WisemanMilburn2009,Jacobs2014}
\begin{align}
\label{eq:kalman_mean}
d\bar{\xvec}_\cond(t)
&=
\matF\bar{\xvec}_\cond(t)dt
+\bvec u(t)dt
+\cov_\cond(t)\matH^\transpose d\wvec(t),\\
\label{eq:kalman_cov}
\dot{\cov}_\cond(t)
&=
\matF\cov_\cond(t) +\cov_\cond(t)\matF^\transpose +\matD -\cov_\cond(t)\matH^\transpose\matH\cov_\cond(t).
\end{align}
Here, $\matF$, $\matD$, and $\matH$ are the drift, diffusion, and measurement matrices, respectively, while $\bvec$ specifies the control channel and $u(t)$ the applied feedback signal. The Wiener increment
\begin{align}
d\wvec(t)=d\yvec(t)-\matH\bar{\xvec}_\cond(t)dt
\end{align}
is the innovation associated with the measurement record.

We consider linear feedback based on the conditional mean, of the form
\begin{align}
u(t)=-\matK\bar{\xvec}_\cond(t),
\label{feedback_law}
\end{align}
where the gain matrix $\matK$ is chosen to optimize the steady-state dynamics, as detailed below. In the long-time limit, the conditional covariance approaches the stationary solution of Eq.~\eqref{eq:kalman_cov}, while the conditional mean remains stochastic and fluctuates according to the measurement record. These fluctuations are central to the certification protocol, since their stationary covariance and temporal correlations determine both the unconditional state and the rate at which independent information is acquired.

\subsection{Excess covariance and unconditional state}
\label{app:excess_covariance}

In addition to the conditional covariance $\cov_\cond(t)$, the conditional mean $\bar{\xvec}_\cond(t)$ fluctuates across measurement trajectories. These fluctuations give rise to the excess covariance
\begin{align}
\cov_{\excess,ij}(t)
=
\av{
\left(\bar{\xvec}_{\cond,i}(t)-\av{\hat{\xvec}_i}_\uncond\right)
\left(\bar{\xvec}_{\cond,j}(t)-\av{\hat{\xvec}_j}_\uncond\right)
}_{\rm tr}.
\end{align}
Using the feedback law in Eq.~\eqref{feedback_law}, its evolution is governed by
\begin{align}
\label{eq:excess_covariance}
\dot{\cov}_\excess(t)
=
\drifteff\cov_\excess(t)
+\cov_\excess(t)\drifteff^\transpose
+\cov_\cond(t)\matH^\transpose \matH\cov_\cond(t),
\end{align}
where
\begin{align}
\drifteff=\matF-\bvec\matK
\end{align}
is the effective closed-loop drift matrix. The unconditional covariance then decomposes as
\begin{align}
\cov_\uncond(t)=\cov_\cond(t)+\cov_\excess(t).
\label{eq:unconditional_covariance}
\end{align}
This decomposition has a direct operational interpretation. The excess covariance $\cov_\excess(t)$ can be estimated from the fluctuations of the conditional-mean trajectory $\bar{\xvec}_\cond(t)$, whereas $\cov_\cond(t)$ is inferred from the dynamical model through Eq.~\eqref{eq:kalman_cov}. The unconditional covariance therefore combines an experimentally sampled contribution with a model-inferred one.

We focus on the stationary regime, in which both $\cov_\cond(t)$ and $\cov_\excess(t)$ converge to time-independent values. In the following, $\cov_\cond$ and $\cov_\excess$ denote these stationary solutions, obtained by setting the time derivatives in Eqs.~\eqref{eq:kalman_cov} and~\eqref{eq:excess_covariance} to zero. We further take the unconditional first moments to vanish, $\av{\hat{\xvec}}_\uncond=0$, without loss of generality.

\subsection{Dynamical model and feedback}
\label{app:cm_dynamics}

We now specify the dynamical model for the symmetric two-particle configuration considered in the main text. With the ordering $\xvec=(\pos_\sysa,\mom_\sysa,\pos_\sysb,\mom_\sysb)^\transpose$ and including a mechanical damping rate $\gamma$, the drift matrix associated with the Hamiltonian in Eq.~\eqref{eq:cm_hamiltonian} is
\begin{align}
\matF=
\begin{pmatrix}
0 & \trapfreq & 0 & 0\\
-\trapfreq-2g & -\gamma & 2g & 0\\
0 & 0 & 0 & \trapfreq\\
2g & 0 & -\trapfreq-2g & -\gamma
\end{pmatrix}.
\end{align}
Measurement backaction induces momentum diffusion at a rate $\phr_\backaction$,
\begin{align}
\matD=4(\phr_\backaction+\phr_\dark) \mathrm{diag}(0,1,0,1),
\end{align}
while independent continuous position measurements of the two particles are described by
\begin{align}
\matH=\sqrt{4\Gamma_\meas}
\begin{pmatrix}
1 & 0 & 0 & 0\\
0 & 0 & 1 & 0
\end{pmatrix},
\end{align}
where $\Gamma_\meas$ is the measurement rate. We define the effective measurement efficiency as
\begin{align}
\eta=\frac{\Gamma_\meas}{\Gamma_\backaction+\Gamma_\dark}.
\end{align}
Introducing the detection efficiency $\eta_\detection=\Gamma_\meas/\Gamma_\backaction$ used in the main text, this can equivalently be written as
\begin{align}
\eta=\frac{\eta_\detection\Gamma_\backaction}
{\Gamma_\backaction+\Gamma_\dark}.
\end{align}
with $\eta=1$ for $\phr_\dark=0$ and $\eta_\detection=1$, and $\eta<1$ in the presence of unmonitored dark diffusion or imperfect detection. In the long-record regime considered for the release--recapture readout, the corresponding isotropic added noise is $\measinp \simeq\eta^{-1/2}$, yielding
\begin{align}
\measinp \simeq
\sqrt{\frac{\Gamma_\backaction+\Gamma_\dark}
{\eta_\detection\Gamma_\backaction}},
\end{align}
as used in the main text.

To stabilize the dynamics and enhance the steady-state entanglement, we employ linear--quadratic--Gaussian (LQG) feedback control; see App.~\ref{app:details_lqg} for details. Following Ref.~\cite{Winkler2025}, we choose a quadratic state cost that favors EPR correlations,
\begin{align}
\matP_{\rm EPR}
=
\trapfreq
\begin{pmatrix}
1 & 0 & -1 & 0\\
0 & 1 & 0 & 1\\
-1 & 0 & 1 & 0\\
0 & 1 & 0 & 1
\end{pmatrix},
\end{align}
thereby penalizing fluctuations of the EPR quadratures $\posop_\sysa-\posop_\sysb$ and $\momop_\sysa+\momop_\sysb$. This choice is appropriate for $g>0$.

We consider a single collective control channel generated by an electric field $E_{\rm fb}(t)$ acting on particles with charges $Q_\sysa$ and $Q_\sysb$. We parameterize the control input as
\begin{align}
u(t)
=
\trapfreq
\frac{Q_\sysa+Q_\sysb}{\sqrt{2}}
\frac{E_{\rm fb}(t)}{\trapfreq\zpmom},
\end{align}
with control vector
\begin{align}
\bvec^\transpose = \begin{pmatrix}
0 & \dfrac{\sqrt{2}Q_\sysa}{Q_\sysa+Q_\sysb} & 0 & \dfrac{\sqrt{2}Q_\sysb}{Q_\sysa+Q_\sysb}
\end{pmatrix}.
\end{align}
We assume $Q_A+Q_B\neq0$.
Controllability of the two-mode dynamics requires $Q_\sysa\neq Q_\sysb$, since for equal charges the electric field couples only to the common mechanical mode. In the main text, we quote an actuation ratio $5:1$, by which we mean that the second component of $\bvec$ is $5$ times larger than the fourth. Note this is different than the configuration $(+e,-e)$ used to estimate the Coulomb force reference value of the main text. The control effort is penalized by a quadratic cost with weight $q$, where $q>0$ controls the strength of the feedback: increasing $q$ suppresses the applied force, whereas decreasing it allows stronger control. Together with the feedback law in Eq.~\eqref{feedback_law}, these choices fully specify the LQG control problem.

\subsection{Linear--quadratic--Gaussian (LQG) feedback control}
\label{app:details_lqg}

In this appendix, we summarize the LQG feedback scheme used in the continuously monitored protocol. Throughout this work, we restrict to linear feedback of the form
\begin{align}
    u(t) = -\matK \bar{\xvec}_\cond(t),
\end{align}
where $\bar{\xvec}_\cond(t)$ denotes the conditional mean vector obtained from the Kalman filter and $\matK$ is the feedback gain matrix.

The feedback is chosen by minimizing the quadratic cost functional~\cite{Jacobs2014}
\begin{align}
\label{eq:lqg_cost}
    J = \lim_{T\to\infty} \frac{1}{T} \left\langle\int_0^T dt \left[\bar{\xvec}_\cond^\transpose(t)\matP\bar{\xvec}_\cond(t) + q u^2(t)\right]\right\rangle_{\rm tr},
\end{align}
where $\matP$ penalizes fluctuations of the mechanical quadratures and $q>0$ penalizes the control effort, with $\trapfreq q$ defining the corresponding dimensionless control-cost parameter. In the main text, we choose $\matP=\matP_{\rm EPR}$ in order to favor Einstein--Podolsky--Rosen-type correlations between the two particles~\cite{Winkler2025}.

For linear dynamics and Gaussian noise, the optimal stationary controller is obtained from the algebraic Riccati equation
\begin{align}
\label{eq:lqr_riccati}
    \matF^\transpose \matY + \matY\matF - \matY\bvec q^{-1}\bvec^\transpose\matY + \matP = 0,
\end{align}
where $\matY$ is the positive semidefinite Riccati matrix. The corresponding optimal feedback gain matrix is
\begin{align}
\label{eq:lqg_gain}
    \matK = q^{-1}\bvec^\transpose\matY.
\end{align}

The resulting closed-loop dynamics are generated by the effective drift matrix
\begin{align}
    \drifteff = \matF - \bvec\matK.
\end{align}
Provided the pair $(\matF,\bvec)$ is controllable and $(\matF,\matP^{1/2})$ is detectable, Eqs.~\eqref{eq:lqr_riccati} and~\eqref{eq:lqg_gain} define a unique stabilizing feedback controller~\cite{WisemanMilburn2009,Jacobs2014}.

\subsection{Covariance estimation from continuous records}
\label{subsec:cm_cov_estimate}

The unconditional steady-state covariance $\cov_\uncond$ is not sampled directly from the continuous measurement record. As discussed above, its excess contribution $\cov_\excess$ is instead reconstructed from the stationary fluctuations of the conditional mean. Let $\bar{\psvec}_\cond(t)$ denote the zero-mean conditional phase-space trajectory. Its stationary autocorrelation matrix is
\begin{align}
\matR_{ij}(\tau)
=
\av{\bar{\psvec}_{\cond,i}(t+\tau)\bar{\psvec}_{\cond,j}(t)}_{\rm tr},
\label{eq:cm_autocorrelation}
\end{align}
with $\matR(0)=\cov_\excess$.

In the stationary regime, $\bar{\psvec}_\cond(t)$ follows a Gaussian Ornstein--Uhlenbeck process generated by Eq.~\eqref{eq:kalman_mean}. Assuming ergodicity, ensemble averages can be obtained from a sufficiently long trajectory,
\begin{align}
\cov_{\excess,ij}
=
\av{\bar{\psvec}_{\cond,i}(t)\bar{\psvec}_{\cond,j}(t)}_{\rm tr}
=
\lim_{T\rightarrow\infty}\frac{1}{T}\int_0^Tdt \bar{\psvec}_{\cond,i}(t)\bar{\psvec}_{\cond,j}(t).
\end{align}
For a finite observation time $T$, we therefore define the estimator
\begin{align}
\estimate{\cov}_{\excess,ij}
=
\frac{1}{T}\int_0^Tdt \bar{\psvec}_{\cond,i}(t)\bar{\psvec}_{\cond,j}(t),
\label{eq:excess_covariance_estimator}
\end{align}
which satisfies $\av{\estimate{\cov}_\excess}_{\rm tr}=\cov_\excess$ in the stationary regime.

To determine its statistical uncertainty, we exploit the Gaussian and stationary character of the process. For observation times much longer than the correlation time, the asymptotic covariance of the estimator is given by Bartlett's formula~\cite{Bartlett1946, BrockwellDavis1991},
\begin{align}
\label{eq:bartlett_covariance}
\Cov\!\left(\estimate{\cov}_{\excess,ij},\estimate{\cov}_{\excess,kl}\right)
\simeq
\frac{1}{T}\int_{-\infty}^{\infty}d\tau 
\left[
\matR_{ik}(\tau)\matR_{jl}(\tau)
+
\matR_{il}(\tau)\matR_{jk}(\tau)
\right].
\end{align}
Equation~\eqref{eq:bartlett_covariance} is the continuous-time analogue of the Wishart covariance. The factor $1/\nm$ associated with independent samples is replaced by the integrated temporal correlations divided by the observation time $T$. We next propagate these fluctuations to the entanglement witness and identify the effective number of independent measurements contained in a continuous record.

\subsection{Entanglement certification in continuous time}
\label{subsec:continuous_correlations}

We consider the witness $\witness=\sse=f(\cov_\uncond)$, with $\cov_\uncond=\cov_\cond+\cov_\excess$. The conditional covariance $\cov_\cond$ is obtained from the calibrated dynamical model, while $\cov_\excess$ is estimated from the stationary, zero-mean record of conditional means. The fluctuations of this covariance estimator obey the Bartlett formula, Eq.~\eqref{eq:bartlett_covariance}. Applying the delta method gives
\begin{align}
    \Var[\estimate{\witness}]
    \simeq \frac{2}{T}\int_{-\infty}^{\infty}d\tau\,
    \trc[\matG(\cov_\uncond)\matR(\tau)\matG(\cov_\uncond)\matR^\transpose(\tau)],
    \label{eq:cm_sse_variance_general}
\end{align}
where $\matR(0)=\cov_\excess$. This expression assumes a sufficiently long stationary record, integrable correlations, and a differentiable witness. Calibration uncertainty in $\cov_\cond$ is not included.

To separate fluctuation amplitude from temporal correlations, we define, in analogy with the release--recapture case,
\begin{align}
    \beta^2
    \equiv 2\trc[\matG(\cov_\uncond)\cov_\excess\matG(\cov_\uncond)\cov_\excess].
    \label{eq:cm_fluctuation_coefficient}
\end{align}
This is the asymptotic variance coefficient for independent samples of the conditional mean. It involves $\cov_\excess$, rather than $\cov_\uncond$, because only the former is estimated from the record. For $\beta>0$, we define the effective correlation time
\begin{align}
    \corrtime\equiv\frac{2}{\beta^2}\int_{-\infty}^{\infty}d\tau\,
    \trc[\matG(\cov_\uncond)\matR(\tau)\matG(\cov_\uncond)\matR^\transpose(\tau)].
    \label{eq:effective_correlation_time}
\end{align}
The asymptotic fluctuations then take the independent-sampling form
\begin{align}
    \Std[\estimate{\tilde{\nu}}_-]
    \simeq\frac{\beta}{\sqrt{T/\corrtime}},
    \label{eq:cm_sse_variance}
\end{align}
with $T/\corrtime$ playing the role of $\nm$. Thus, $\beta$ measures the fluctuation amplitude per independent sample, while $\corrtime$ determines the rate at which statistically equivalent information is acquired. Both depend on the state and the measurement and feedback dynamics.

The interpretation of $\corrtime$ is explicit for exponentially decaying correlations,
\begin{align}
    \matR(\tau)=\cov_\excess e^{-|\tau|/\bar{\corrtime}}.
\end{align}
The integrand in Eq.~\eqref{eq:effective_correlation_time} then decays as $\exp(-2|\tau|/\bar{\corrtime})$, yielding
\begin{align}
    \corrtime
    =\int_{-\infty}^{\infty}e^{-2|\tau|/\bar{\corrtime}}\,d\tau
    =\bar{\corrtime}.
    \label{eq:corrtime_toy_model}
\end{align}
For general dynamics, $\corrtime$ weights temporal correlations according to their contribution to the witness fluctuations and need not coincide with the decay time of any individual quadrature.

With these definitions, and for $\sse<1$, the $\nsigma$-sigma certification time is
\begin{align}
    T_\star\simeq\left(\frac{\nsigma\beta}{1-\sse}\right)^2\corrtime,
    \label{eq:cm_certification_time}
\end{align}
which has the same structure as the release--recapture result: the sample requirement is multiplied by the time needed to acquire an independent sample. In continuous monitoring, this time is $\corrtime$, whereas in release--recapture it is the cycle duration $\tint+\trest$. This expression describes stationary acquisition; any initial preparation time must be added separately.

\bibliography{bibliography}